\documentclass[12pt,journal,draftcls,a4paper,onecolumn]{IEEEtran}

\usepackage{dsfont}
\usepackage{amsmath}
\usepackage{amssymb}
\usepackage{amsfonts}
\usepackage{graphicx}
\usepackage{amsmath}
\usepackage[all,poly]{xy}
\usepackage{multirow}
\usepackage{algorithm}
\usepackage{algorithmic}
\usepackage{color}
\usepackage[noadjust]{cite}
\usepackage{subfigure}
\usepackage{tikz,pgf}
\usetikzlibrary{fit}
\usetikzlibrary{arrows,automata}
\usepackage{verbatim}
\usetikzlibrary{positioning,shapes.geometric}
\usepackage{url}
\usepackage[T1]{fontenc}

\newcommand{\figwidth}{\columnwidth}

\newcommand{\bzero}{\boldsymbol{0}}

\newcommand{\bSig}{\boldsymbol{\Sigma}}

\newcommand{\bLam}{\boldsymbol{\Lambda}}
\newcommand{\beps}{\boldsymbol{\epsilon}}

\def\bsa{{\boldsymbol{a}}}
\def\bsb{{\boldsymbol{b}}}

\def\bse{{\boldsymbol{e}}}

\def\bsm{{\boldsymbol{m}}}

\def\bsv{{\boldsymbol{v}}}
\def\bsw{{\boldsymbol{w}}}

\def\bsy{{\boldsymbol{y}}}

\def\bsA{{\boldsymbol{A}}}

\def\bsE{{\boldsymbol{E}}}

\def\bsI{{\boldsymbol{I}}}

\def\bsM{{\boldsymbol{M}}}

\def\bsR{{\boldsymbol{R}}}
\def\bsS{{\boldsymbol{S}}}

\def\bsV{{\boldsymbol{V}}}
\def\bsW{{\boldsymbol{W}}}

\def\bsY{{\boldsymbol{Y}}}

\def\calN{{\mathcal{N}}}

\newcounter{algo}
\renewcommand{\thealgo}{\arabic{algo}}

\title{Estimating the Intrinsic Dimension of Hyperspectral Images Using an Eigen-Gap Approach}

\author{Abderrahim Halimi$^{\,1}$, Paul Honeine$^{\,1}$, Malika Kharouf$^{\,1}$\thanks{(1) A. Halimi, P. Honeine,  and M. Kharouf  are with the Institut Charles Delaunay (CNRS), Universit\' e de technologie de Troyes, France}, C\'edric Richard$^{\,2}$\thanks{(2) C. Richard is with the University of Nice Sophia-Antipolis, CNRS, Observatoire de la C\^ote d'Azur, France}, Jean-Yves Tourneret$^{\,3}$\thanks{(3) J.-Y. Tourneret is with the University of Toulouse, IRIT/INP-ENSEEIHT/T\'eSA, Toulouse, France}
\thanks{This work was supported in part by the HYPANEMA ANR Project under Grant ANR-12-BS03-003, and in part by the Thematic Trimester on Image Processing of the CIMI Labex, Toulouse, France, under Grant ANR-11-LABX-0040-CIMI within the Program ANR-11-IDEX-0002-02.}}

\begin{document}
\maketitle
\begin{abstract}
Linear mixture models are commonly used to represent hyperspectral datacube as a linear combinations of endmember spectra. However, determining of the number of endmembers for images embedded in noise is a crucial task. This paper proposes a fully automatic approach for estimating the number of endmembers in hyperspectral images. The estimation is based on recent results of random matrix theory related to the so-called spiked population model. More precisely, we study the gap between successive eigenvalues of the sample covariance matrix constructed from high dimensional noisy samples. The resulting estimation strategy is unsupervised and robust to correlated noise. This strategy is validated on both synthetic and real images. The experimental results are very promising and show the accuracy of this algorithm with respect to state-of-the-art algorithms.
\end{abstract}
\section{Introduction}
\label{sec:Introduction}

%
%
%

Unmixing techniques can provide fundamental information when analyzing multispectral or hyperspectral images with limited spatial resolution. In spite of almost 50 years of research in this area, there has been a surge of interest in the last few years within the area of remote sensing and hyperspectral imaging  \cite{Keshava2002,Bioucas2012}. Even with an ever-increasing spatial resolution, each pixel (or spectrum) in a hyperspectral image is generally associated with several pure materials. Each spectrum can thus be seen as a mixture of spectral signatures  called endmembers with respective proportions called abundances. While non-linear unmixing techniques have been recently investigated \cite{Dobigeon2014,Halimi2011TGRS,AltmannTIP2011}, the linear mixing model is widely accepted because of its natural physical interpretation. This model assumes that each spectrum is a convex combination of the endmember spectra. Unmixing hyperspectral images consists of three stages: (i) determining the number of endmembers and possibly projecting the data onto a subspace of reduced dimension \cite{Bioucas2008,ChangTGRS2004}, (ii) extracting endmember spectra \cite{Nascimento2005,Dobigeon2009} and (iii) estimating their abundances \cite{Heinz2001,Chen2014,NguyenEAS2013}. These stages can be performed separately or jointly \cite{Ammanouil2014,HalimiArxiv2014,HoneineTGRS2012}. Determining the number of endmembers, or the signal subspace dimension, then appears  as a fundamental step in order to achieve  endmember spectrum determination and abundance estimation. This paper considers this problem of estimating the signal subspace dimension of hyperspectral images.


Estimating the number of endmembers present in a scene has been described under several names and under different methodological frameworks. The most well known definitions are based on the eigenvalues of the sample (observation) covariance matrix, with the so-called ``virtual dimension'' (VD), as well as many variants including the ``intrinsic dimension'' and the ``effective dimension''. The VD is estimated by the so-called Harsanyi-Farrand-Chang (HFC) method which relies on the Neyman-Pearson detection theory applied to the difference between the eigenvalues of the sample covariance matrix and its non centered counterpart (i.e., the matrix of second-order moments) \cite{ChangTGRS2004}. The HFC, and its noise whitened version (NWHFC), are generally more efficient than algorithms based on model selection criteria such as the Akaike information criterion (AIC) \cite{Akaike1974} and the minimum description length (MDL) \cite{schwarz1978,Rissanen1978}, especially in the presence of colored noise. The idea of evaluating the differences between the eigenvalues of the covariance  and the correlation  matrices has also been exploited in other algorithms such as  \cite{Luo2013}.
In \cite{Bioucas2008}, the authors proposed an unsupervised approach for hyperspectral subspace identification called Hysime. Their method consists of minimizing a cost function whose aim is to reduce the noise power. Other methods only use the sample covariance matrix without considering the correlation matrix. In \cite{ChangTGRS2004}, the noise subspace projection method considers a Neyman-Pearson test to separate  noise components from signal components based on a whitened covariance matrix.  The idea is that the noise eigenvalues are equal to unity while the signal eigenvalues are greater than one.	

Random matrix theory (RMT) is a universal multivariate statistics tool that has been used successfully in many fields.
Recently, an approach based on RMT has been applied to estimate the number of endmembers \cite{CawseTIP2013}. This method (denoted as RMT) first estimates  the noise covariance matrix in order to remove colored noise effects. Then, based on the whitened covariance eigenvalues, this method proposes a theoretical threshold to determine the number of endmembers in the image. However, this method appears to be sensitive to noise \cite{CawseWHISPERS2011,CawseTIP2013} which might reduce the estimation performance. Moreover, it has been shown in \cite{CawseWHISPERS2012} that many noise estimation algorithms are sensitive to noise correlation. In the presence of such correlations, the RMT algorithm \cite{CawseTIP2013} may provide results of poor quality.

The motivation of our work is to provide a consistent and unsupervised estimator of the number of endmembers by considering a general scenario, where the additive noise components are not identically distributed. The main advantage of the proposed approach, with respect to (w.r.t.) the RMT algorithm, is its robustness in the presence of correlated noise. Similarly to the RMT and Hysime approaches, our method starts by estimating the noise covariance matrix in order to remove its effect from the sample/observation covariance matrix. The next step is inspired from recent results on spiked population models (SPM). Indeed, \cite{Passemier_RMTA12} proposed a method based on the   gap between successive eigenvalues of the sample covariance matrix. By considering sorted   eigenvalues, the main idea is that the gap between eigenvalues (of a whitened covariance) is larger in the presence of a signal while it is reduced for noise components. Building on this idea, an automatic threshold is obtained to separate the signal from noise components. 

\subsection*{Contributions and comparisons} \label{sec:Contributions}
The main objective of the paper is to provide an unsupervised algorithm for estimating the number of endmembers in hyperspectral images.
The proposed approach generalizes the consistent estimator proposed in \cite{Passemier_RMTA12} for independent and identically distributed (i.i.d.) noise to the colored Gaussian noise case. This eigen-gap approach is based on a consistent estimator while the RMT algorithm \cite{CawseTIP2013} is not fully consistent as stated in \cite{Passemier_RMTA12}. The proposed approach appears to be more robust to correlated noise and to small image sizes.  These statements are validated on both synthetic and real hyperspectral images.

The paper is organized as follows. Section \ref{sec:Problem_formulation} introduces the hyperspectral mixing model, the SPM and rank estimation methods for an SPM. Section \ref{sec:Algorithm} introduces our algorithm whose performance is evaluated in Section  \ref{sec:Simulation_results_on_synthetic_data} on synthetic images. Results on real hyperspectral images are presented in Section \ref{sec:Simulation_results_on_real_data}. Conclusions and perspectives for future works are finally reported in Section
\ref{sec:Conclusions}.


%
%
%

\section{Problem formulation}
\label{sec:Problem_formulation}

\subsection{Linear mixture model}
\label{subsec:Linear_mixture_model}
The linear mixture model (LMM) assumes that each pixel spectrum $\bsy_n$, of size $L\times 1$,
is a linear combination of $R$ endmembers $\bsm_{r}$, $r\in \left\lbrace 1,\cdots,R \right\rbrace$, corrupted by an additive   noise $\bse_{n}$ as follows
\begin{eqnarray}
\bsy_n & = & \sum_{r=1}^{R}{a_{rn} \bsm_{r}}+ \bse_{n}  \nonumber \\
       & = &  \bsM \bsa_n +\bse_{n}   \label{eqt:LMM_1pixel}
\end{eqnarray}
with $\bse_{n} \sim \calN \left(\boldsymbol{0}_L, \bSig\right)$ a Gaussian noise,  $\bSig $ is the noise covariance matrix,   $\boldsymbol{0}_L$ is an  $L \times 1$ vector of $0$,  $\bsa_n = \left[a_{1n}, \cdots, a_{Rn} \right]^{\top}$ is the $R \times 1$ abundance vector of the $n$th pixel and $\bsM = \left[\bsm_1,\cdots,\bsm_R\right]$ is an $L \times R$ matrix gathering the endmember spectra. The abundance vector $\bsa_n$ contains proportions satisfying the  positivity and sum-to-one (PSTO) constraints $a_{rn} \geq 0, \forall r \in \left\{1,\ldots,R\right\} \quad \textrm{and} \quad \sum_{r=1}^{R}{a_{rn}}=1.$
Considering  $N$ pixels gathered in the $L\times N$ matrix $\bsY$, the LMM can be written as follows
\begin{equation}
\bsY =  \bsM \bsA +\bsE  \label{eqt:LMM_all}
\end{equation}
where $\bsA$ is an $R \times N$ matrix of abundances, and $\bsE$   an $L \times N$ matrix of noise samples.

Rank estimation can be based on an eigen-value analysis of the covariance matrix of $\bsY$. Assuming independence between the signal counterpart $\bsS = \bsM \bsA$ and the noise  $\bsE$ leads to 
\begin{equation}
\bsR_{Y} =   \bsR_{S} + \bSig   \label{eqt:Cov_LMM}
\end{equation}
where $\bsR_{Y}$ and $\bsR_{S}$ are the covariance matrices of $\bsY$ and $\bsS$, respectively. In this paper, we are interested in estimating the number  $R$ of endmembers, which is equal to   $K+1$, where $K=\textrm{rank}(\bsR_{S})$. Indeed, the signal lies into a subspace of dimension $R-1$ because of the PSTO constraints.

\subsection{Spiked population model}
\label{subsec:Spiked population model}
A well-known model in RMT is the spiked population model. This model assumes that the covariance matrix of interest has all its eigenvalues equal to  $\sigma^2$ except  a few  eigenvalues (known as spikes) as follows \cite{Passemier_RMTA12}
\begin{equation}
\bLam =  \sigma^2 \boldsymbol{\Gamma} \left[ \begin{array}{ccc|c}
   \gamma_1 &          &           &  \\
            &  \ddots  &           & \bzero_{K,L-K} \\
            &          &  \gamma_K & \\     \hline
 & \bzero_{L-K,K}  &           &  \bsI_{L-K}
\end{array} \right] \boldsymbol{\Gamma}^{\top}    \label{eqt:Cov_SPM}
\end{equation}
where $\bLam$ is the covariance matrix, $\boldsymbol{\Gamma}$ is an $L \times L$ orthogonal  matrix, $\bzero_{i,j}$ is the $i \times j$ matrix of $0$ and $ \bsI_L$ is the $L \times L$ identity matrix.
Determining the number of  endmembers can be performed by computing the number of spiked eigenvalues of the covariance matrix $\bLam$. For this purpose, consider that $\bsR_{Y}=\bLam$ and denote its eigenvalues  by $\lambda_k $ for $k= 1,\cdots,L$. By assuming\footnote{In presence of colored noise, an adequate procedure will be considered as shown in the following.} $\bSig  = \sigma^2 \bsI_L$   and the  eigenvalue vector $\left[\rho_1, \cdots,\rho_K, \bzero_{1,L-K}\right]^{\top}$ for $\bsR_S$,  \eqref{eqt:Cov_LMM}   leads to
\begin{equation}
\rho_k + \sigma^2 = \gamma_k  \sigma^2, \textrm{ for  }  k \leq K \label{eqt:eigen_Y2}
\end{equation}
and \eqref{eqt:Cov_SPM} yields
\begin{equation}
\lambda_k = \left\lbrace
\begin{array}{lcl}
\rho_k + \sigma^2,  &\textrm{ if  }  k \leq K    \\
\sigma^2, &\textrm{ otherwise.  }
\end{array}  \right.
\label{eqt:eigen_Y}
\end{equation}
Unfortunately, in many situations, the  covariance matrix $\bsR_S$ is unknown and the additive noise is not necessarily identically distributed contradicting the assumption $\bSig  = \sigma^2 \bsI_L$. The alternative proposed in this work builds an estimator of the number of endmembers, when only the sample covariance matrix $\bsR_{Y}$ is known and $\bse_n$ is an additive independently and not identically distributed zero-mean Gaussian noise sequence.

\subsection{Rank estimation from an SPM}
\label{subsec:Rank_estimation_from_an_SPM}
Estimating the number of spikes from an SPM is an interesting problem that has found many applications including chemical mixtures \cite{KritchmanTSP2009} and hyperspectral unmixing \cite{CawseTIP2013}. A recent work proposed to investigate RMT to estimate the number of spikes or endmembers  in  hyperspectral images  \cite{CawseTIP2013}. This work builds on the estimator proposed in \cite{KritchmanTSP2009} in the context of chemical mixtures. This method uses the following assumptions: (i) $N \rightarrow \infty$ and $L \rightarrow \infty$ (or large values of $N$ and $L$) with $c=\frac{L}{N} > 0$ a positive constant ,  (ii) the noise corrupting the data is Gaussian and independent of the signal, (iii) the signal covariance matrix has a fixed rank $K$.
Under these assumptions, the method \cite{CawseTIP2013,KritchmanTSP2009} is based on the study of the asymptotic behavior of the largest eigenvalues of the sample covariance matrix when both the dimension of the observations and the sample size grow to infinity at the same rate. The main idea is that when the covariance matrix $\bLam$ is a perturbed version of a finite rank matrix, all but a finite number of eigenvalues of the covariance matrix are different from the i.i.d. noise variance.
Based on this property and on \cite{Johnstone2001,Baik_JMA2006}, a threshold that separates the eigenvalues corresponding to the useful information from those corresponding to the noise was derived in \cite{CawseTIP2013,KritchmanTSP2009} yielding
\begin{equation}
\widehat{K} = \min_{k=1,\cdots,L}\left( \lambda_k < \sigma^2 \left(\frac{\beta_{c}}{N^{2/3}} s(\alpha) + (1+\sqrt{c})^2\right)\right) - 1
\end{equation}
where $\lambda_1 \ge \lambda_2 \ge \dots \ge \lambda_L$ are the eigenvalues of the sample covariance matrix $\bLam$, $s(\alpha)$ can be found by using the Tracy-Widom distribution, and
\begin{equation}
\beta_{c} = \left(1+ \sqrt{c}\right)\left(1+ \sqrt{c^{-1}}\right)^{1/3}.
\end{equation}
This estimator is based on a sequence of nested hypothesis tests. By construction, the proposed estimator is not fully consistent as shown in \cite{PassemierPHD2012}. In this paper, we are interested in deriving a new estimator with better statistical properties.

One of the front-line research problems in RMT is the study of the gap between consecutive eigenvalues \cite{Baik_JMA2006,Passemier_RMTA12,PassemierPHD2012}. Indeed, the eigenvalue differences can be used for the estimation of the number of spikes under the following assumptions \cite{Passemier_RMTA12}: (i)  $N$ and $L$ are related by the asymptotic regime $N \rightarrow \infty,$ $\frac{L}{N} \rightarrow c > 0$,  (ii) the noise corrupting the data is Gaussian and independent of the signal (to satisfy assumption 3.1 in \cite{Passemier_RMTA12}), (iii) the signal covariance matrix has a fixed rank $K$, (iv) the eigenvalues of the sample covariance matrix are of multiplicity one\footnote{The general case of multiple multiplicity has been considered in \cite{PassemierPHD2012}.} and (v) $\gamma_1 > \cdots >\gamma_K > 1+\sqrt{c}$. Note first that using hypotheses (i) and (v),  it is shown in \cite{Baik_JMA2006} that the eigenvalues of the covariance matrices of spiked population models satisfy almost surely
\begin{equation}
\lambda_k  \xrightarrow[N\rightarrow \infty ]{a.s.}  \sigma^2 \phi(\gamma_k)   \label{eqt:lambda_conv1}
\end{equation}
for each $k \in \left\lbrace1,\cdots,K \right\rbrace$, while, for $ k > K$
\begin{equation}
\lambda_k \xrightarrow[N\rightarrow \infty ]{a.s.}  \sigma^2  (1 + \sqrt{c})^2  \label{eqt:lambda_conv2}
\end{equation}
where $\phi(x)$ is defined by
\begin{equation}
\phi(x) = \left( x + 1 \right) \; \left( 1+ \frac{c}{x} \right).
\end{equation}
These results were used in \cite{Onatski_Econ08,Passemier_RMTA12} to infer the number of components $K$ in the case where $\sigma^2=1$.
In the general case where $\sigma^2 \neq 1$, the authors of \cite{Passemier_RMTA12} stated that one should divide the eigenvalues by the noise variance $\sigma^2$ to apply the results obtained for $\sigma^2=1$.  The estimation method in \cite{Passemier_RMTA12}  considers  the following differences between successive eigenvalues
\begin{equation}
\delta_k = \lambda_k - \lambda_{k+1}, \textrm{ for  } k=1,\cdots,L-1.
\end{equation}
The main idea is that, when approaching non-spiked values, the  eigen-gap $\delta_k$ shrinks to small values. Therefore, the  number of endmembers can be estimated as follows
\begin{equation}
\widehat{K} =  \mbox{min}\left\{ k \in \{1, \dots , M\};  \delta_{k+1} < d_N\right\}
\end{equation}
where  $M \geq K$ is a fixed integer (large enough), and $d_N \xrightarrow{n \rightarrow \infty} 0$ is a threshold to determine. According to \cite{Passemier_RMTA12}, the consistency of this estimator is ensured if $d_N \rightarrow 0$ and $N^{2/3} \; d_N \rightarrow +\infty$. The same authors proposed to use $d_N = \frac{\psi_N}{N^{2/3}} \beta_{c}$ with $\psi_N = 4 \sqrt{2 \log(\log N)}$ that satisfies the former conditions. The obtained algorithm was fully unsupervised in the sense that it did not require to tune any parameter.
The main difference between this strategy and \cite{CawseTIP2013,KritchmanTSP2009} is that \cite{Passemier_RMTA12}  builds a test statistics based on the gaps between successive eigenvalues and not on the eigenvalues themselves. An important consequence is that a theoretical estimator consistency is ensured in the case of the gap approach while the method described in \cite{CawseTIP2013,KritchmanTSP2009} depends on a parameter $\alpha$ and is nearly consistent as stated in \cite{Passemier_RMTA12}.


\section{Proposed algorithm}
\label{sec:Algorithm}
The eigen-gap strategy assumes the noise to be  i.i.d. which is not true when considering  hyperspectral images \cite{Bioucas2008,Altmann2014b}. Therefore, we propose to use a preliminary step before estimating the number of endmembers.

\subsection{Noise estimation}
\label{subsec:Noise_estimation}
A great effort has been devoted to the noise estimation problem since it is essential for many signal processing applications requiring whitening and/or dimension reduction. Among these algorithms, we distinguish  those assuming spatial homogeneous regions such as the nearest neighbor difference (NND) \cite{Green_TGRS1988}, the geometrical based algorithm \cite{Meer_TPAMI1990}, and algorithms estimating the noise  such as the multiple regression based methods \cite{Bioucas2008,ChangTGRS2004,RogerIGRS1996}. The NND algorithm requires homogenous areas that are not always available in hyperspectral images \cite{Bioucas2008}. The Meer algorithm does not account for noise spectral correlation since it estimates the noise variance for each band separately \cite{CawseJSTARS2013}. This paper considers the multiple regression based method proposed in \cite{Bioucas2008} since it has been studied in many subspace identification algorithms \cite{CawseJSTARS2013,AndreouJSTARS2014} and has shown similar results as the residual method of \cite{ChangTGRS2004}  as stated in \cite{CawseJSTARS2013}. However, the proposed approach is still valid when considering other noise estimation algorithms.

The multiple regression method \cite{Bioucas2008} assumes that the $\ell$th spectral band of each pixel vector is connected to the $L-1$ other bands by a linear model.  More precisely, denoting as $\bsy_{\ell} $ the $N\times 1$ vector containing the pixel elements of the $\ell$th band, and $\bsY_{-\ell}$  the $(L-1)\times N$ matrix obtained by removing the $\ell$th row from the matrix $\bsY$,  we assume that
\begin{equation}
\bsy_{\ell}  =  \bsY_{-\ell}^{\top} \bsb_{\ell} + \beps_{\ell}
\end{equation}
where $\beps_{\ell}$ is the modeling error vector of size $N\times 1$ and  $\bsb_{\ell}$ is the $(L-1) \times 1$ regression vector that is estimated using the least squares estimator \cite{Bioucas2008}
\begin{equation}
\widehat{\bsb}_{\ell} = \left(\bsY_{-\ell} \bsY_{-\ell}^{\top}\right)^{-1} \bsY_{-\ell} \bsy_{\ell}.
\end{equation}
The noise vector is then estimated by $\widehat{\beps}_{\ell} = \bsy_{\ell}- \bsY_{-\ell}^{\top} \widehat{\bsb}_{\ell} $
and its covariance matrix is given by
\begin{equation}
\widehat{\bSig} = \left(\widehat{\beps}_{1},\cdots,\widehat{\beps}_{L}\right)^{\top} \left(\widehat{\beps}_{1},\cdots,\widehat{\beps}_{L}\right)/N.
\label{eq:CovEst_Noise}
\end{equation}
Once the noise covariance matrix has been estimated, a whitening procedure can be performed as described in the next section.

%
%
%

\subsection{Rank estimation}
\label{subsec:Rank_estimation}
Before applying the eigen-gap test, let us first remove the effect of colored noise. This can be achieved by whitening the observed pixels $\bsY$ using the estimated noise covariance matrix $\widehat{\bSig}$. However, it has been shown in \cite{CawseWHISPERS2011,CawseTIP2013} that  this procedure leads to an overestimated subspace dimension $K$ when combined to RMT approaches. Therefore, we will consider the strategy used in \cite{CawseTIP2013}.
Under the assumption that $\bsv_i^{\top} \bsw_i \neq 0, \forall i=1,\cdots,L$,  it has been shown in \cite{CawseTIP2013} that
\begin{equation}
\widehat{\lambda}_k = \left\lbrace
\begin{array}{ll}
\rho_k + \frac{\bsv_k^{\top} \widehat{\bSig} \bsw_k}{\bsv_k^{\top} \bsw_k}, &\textrm{ if  }  k \leq K    \\
\frac{\bsv_k^{\top} \widehat{\bSig} \bsw_k}{\bsv_k^{\top} \bsw_k}, &\textrm{ otherwise  }
\end{array}  \right.   \label{eqt:eigen_whitening}
\end{equation}
where $\bsv_k$ and $\bsw_k$ denote  the eigenvectors of $\bsR_{Y}$ and   $\bsR_{S}$, respectively. Note that \eqref{eqt:eigen_whitening} is similar to \eqref{eqt:eigen_Y} except that the noise variance has been estimated differently as
\begin{equation}
\widehat{\sigma}^2_k = \frac{\bsv_k^{\top} \widehat{\bSig} \bsw_k}{\bsv_k^{\top} \bsw_k} \label{eqt:New_noise}.
\end{equation}
Equations \eqref{eqt:eigen_whitening} and \eqref{eqt:New_noise} require the computation of the eigenvectors of $\bsR_{S}$. The covariance matrix  $\bsR_{S}$ is unknown but can be estimated using \eqref{eqt:Cov_LMM} as follows
\begin{equation}
\widehat{\bsR}_S = \bsR_Y-\widehat{\bSig} \label{eqt:Cov_Signal}.
\end{equation}
Finally,  to account for colored noise, one has to include the noise variance  \eqref{eqt:New_noise} in  \eqref{eqt:lambda_conv1} and \eqref{eqt:lambda_conv2} by dividing each eigenvalue $\lambda_k$ by the corresponding noise variance $\sigma^2_k$,  as stated in \cite{Passemier_RMTA12}. The resulting rank estimator is given by
\begin{equation}
\widehat{K} =  \mbox{min}\left\{ k \in \{1, \dots , M\};  \Delta_{k+1}  < d_N\right\} \label{eqt:K_estimate}
\end{equation}
with
\begin{equation}
\Delta_{k+1} =\frac{\widehat{\lambda}_{k}}{\widehat{\sigma}^2_k } - \frac{\widehat{\lambda}_{k+1}}{\widehat{\sigma}^2_{k+1}}  \textrm{  and  }
d_N = \frac{\psi_N}{N^{2/3}} \beta_{c}.
\label{eqt:Threshold}
\end{equation}
The resulting algorithm is summarized in Algo. \ref{alg:EG_RMT}.
\begin{algorithm}
\caption{Proposed algorithm} \label{alg:EG_RMT}
\begin{algorithmic}[1]
       \STATE Compute the sample covariance matrix $\bsR_Y$
		   \STATE Estimate the noise covariance matrix  $\widehat{\bSig}$
			 \STATE Compute the matrix $\bsV$ containing the eigenvectors of $\bsR_Y$ (sorted in descending order of the eigenvalues)
			 \STATE Compute the matrix  $\bsW$ containing the eigenvectors of $\widehat{\bsR}_S = \bsR_Y-\widehat{\bSig}$ (sorted in descending order of the eigenvalues)
			 \STATE Compute $\widehat{\lambda_k}$, $k \in \left\lbrace 1,\cdots,L\right\rbrace$ the eigenvalues of $\bsR_Y$ (sorted in descending order)
			 \STATE Compute $\widehat{\sigma^2_k}$ according to \eqref{eqt:New_noise}
       \STATE Compute $\Delta_{k+1} $ and $d_N$ according to \eqref{eqt:Threshold}
			 \STATE Estimate the number of endmembers $\widehat{R}=\widehat{K}+1$ by evaluating \eqref{eqt:K_estimate}
\end{algorithmic}
\end{algorithm}

\section{Simulation results on synthetic data} \label{sec:Simulation_results_on_synthetic_data}
\begin{figure}[h!]
\centering
\includegraphics[width=0.91\figwidth]{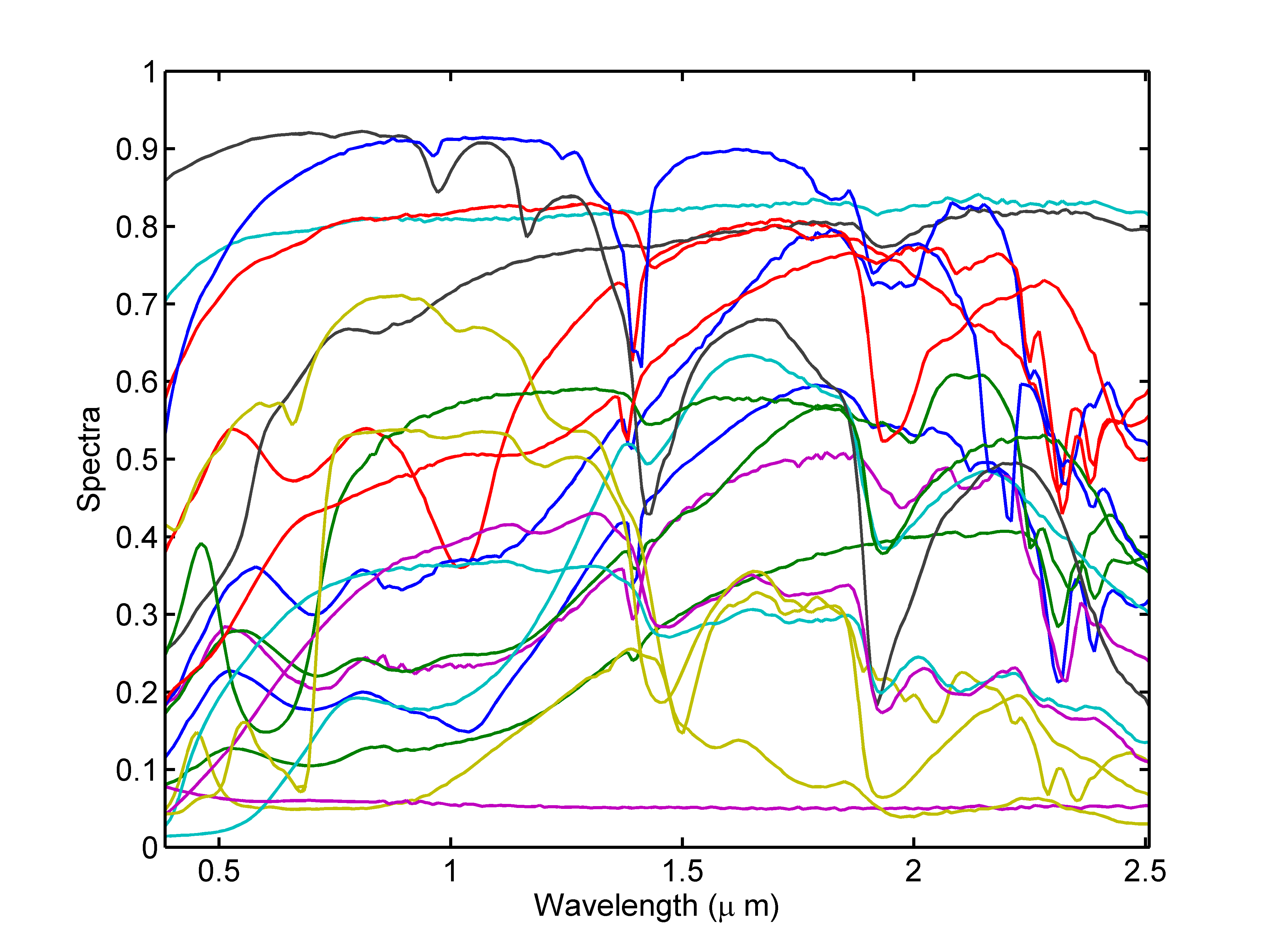}
\caption{Spectra from USGS library.} \label{fig:Exp9_20Spectra}
\end{figure}

This section analyzes the performance of the proposed eigen-gap approach (EGA) with simulated data. The proposed approach is compared to the NWHFC approach since it has been shown in \cite{ChangTGRS2004} to provide better results than the approaches based on information criteria such as  AIC \cite{Akaike1974} and MDL \cite{schwarz1978,Rissanen1978}. Note that the $\textrm{NWHFC}$ algorithm\footnote{The NWHFC is obtained by preceding the HFC algorithm by a noise whitening step. We have considered the  HFC algorithm available in: \url{http://www.ehu.es/computationalintelligence/index.php/Endmember_
Induction_Algorithms}   } requires the definition of the false alarm probability $P_f$. We have considered in our experiments three values $P_f\in \left\lbrace10^{-3},10^{-4},10^{-5}\right\rbrace$  denoted by   $\textrm{NWHFC}_1$, $\textrm{NWHFC}_2$ and  $\textrm{NWHFC}_3$, respectively.
The EGA is also compared to the RMT approach proposed in \cite{CawseTIP2013} since it uses similar theoretical tools. The well known Hysime algorithm \cite{Bioucas2008} is also investigated since it has been used in many studies \cite{CawseTIP2013,AndreouJSTARS2014}. The considered datasets were constructed based on the USGS spectra library used in  \cite{Bioucas2008}. As in \cite{CawseTIP2013}, we considered $20$ minerals that vary widely (some spectra are similar, other are different, some spectra have low amplitude,...) as shown in Fig. \ref{fig:Exp9_20Spectra}.
The  abundances were drawn uniformly in the simplex defined by the PSTO constraints using a Dirichlet distribution \cite{Bioucas2008}.
The following sections present three kinds of results: (i) robustness with respect to noise, (ii) impact of the image size and (iii) performance with respect to the number of endmembers. In all these experiments, we  considered the following parameters $N=10^4$ pixels, $L=224$ bands, $\textrm{SNR}=25$ dB, and $R=4$ endmembers, when fixed according to the experimental setups (i), (ii) or (iii). We performed $50$ Monte-Carlo simulations for each experiment.

\subsection{Robustness to noise} \label{subsec:Robustness_to_noise}
\begin{figure}[h!]
\centering
\includegraphics[width=0.91\figwidth]{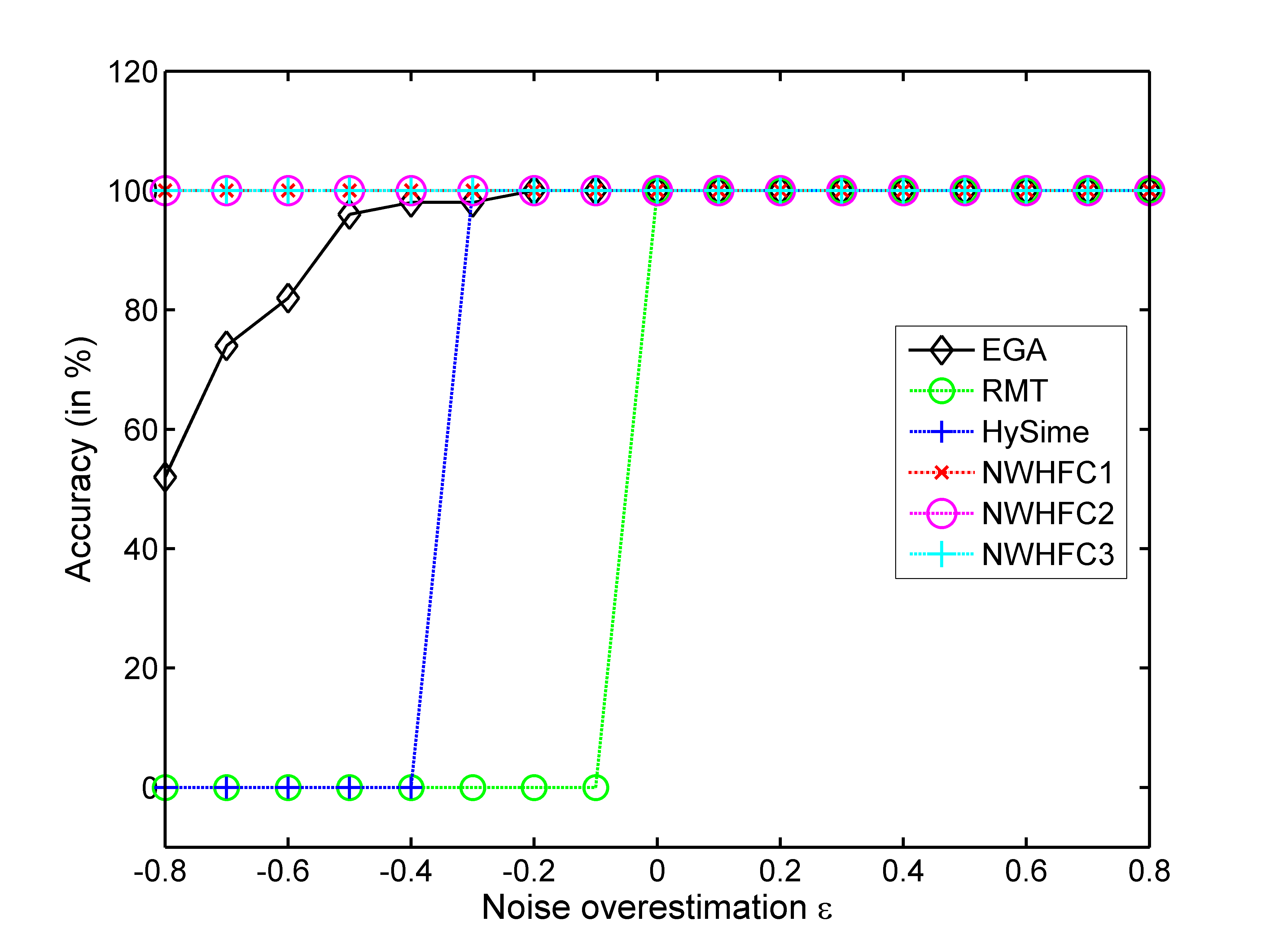}
\caption{Robustness of the algorithms with respect to the accuracy of the noise estimation.} \label{fig:Exp1_Approx_Noise_20Simul}
\end{figure}

This section studies the robustness of the proposed approach with respect to noise. Two experiments were considered.
The first experiment studies the performance of the different algorithms when the noise variance is inaccurately estimated. Indeed, all the algorithms account for a noise estimation step that may introduce some errors. Therefore, we simulated synthetic images using $R=4$ fixed endmembers (chosen from the $20$ spectra) with an i.i.d. Gaussian noise with variance $\sigma^2$ (corresponding to   $\textrm{SNR}=25$ dB). Then, we applied the described algorithms when considering a  noise variance given by $\sigma^2 \left(1+\epsilon\right)$, to simulate an error in the noise estimation step. Fig. \ref{fig:Exp1_Approx_Noise_20Simul} shows the obtained accuracy (in percent) of the estimated  number of endmembers when varying $\epsilon$ (the accuracy represents the percentage of good estimates).
This figure shows the robustness of the algorithms with respect to noise overestimation.  However, observe that both RMT and Hysime algorithms are sensitive to noise variance under-estimation since they provide uncorrect results for $\epsilon\leq-0.1$ and $\epsilon \leq-0.4$, respectively.
The  results show the robustness of the proposed EGA since it provides an accuracy higher than 90$\%$ for $\epsilon>-0.5$. The best performance was obtained with the NWHFC approach. This algorithm  applies a Neyman-Pearson test on the difference between covariance and correlation eigenvalues. Therefore, the additive noise perturbation introduced by $\left(1+\epsilon\right)$ is eliminated (or greatly reduced).  The proposed EGA is more robust than RMT and Hysime to noise estimation errors which is of great interest especially when considering real data.

\begin{figure}[h!]
\centering
\includegraphics[width=0.95\figwidth]{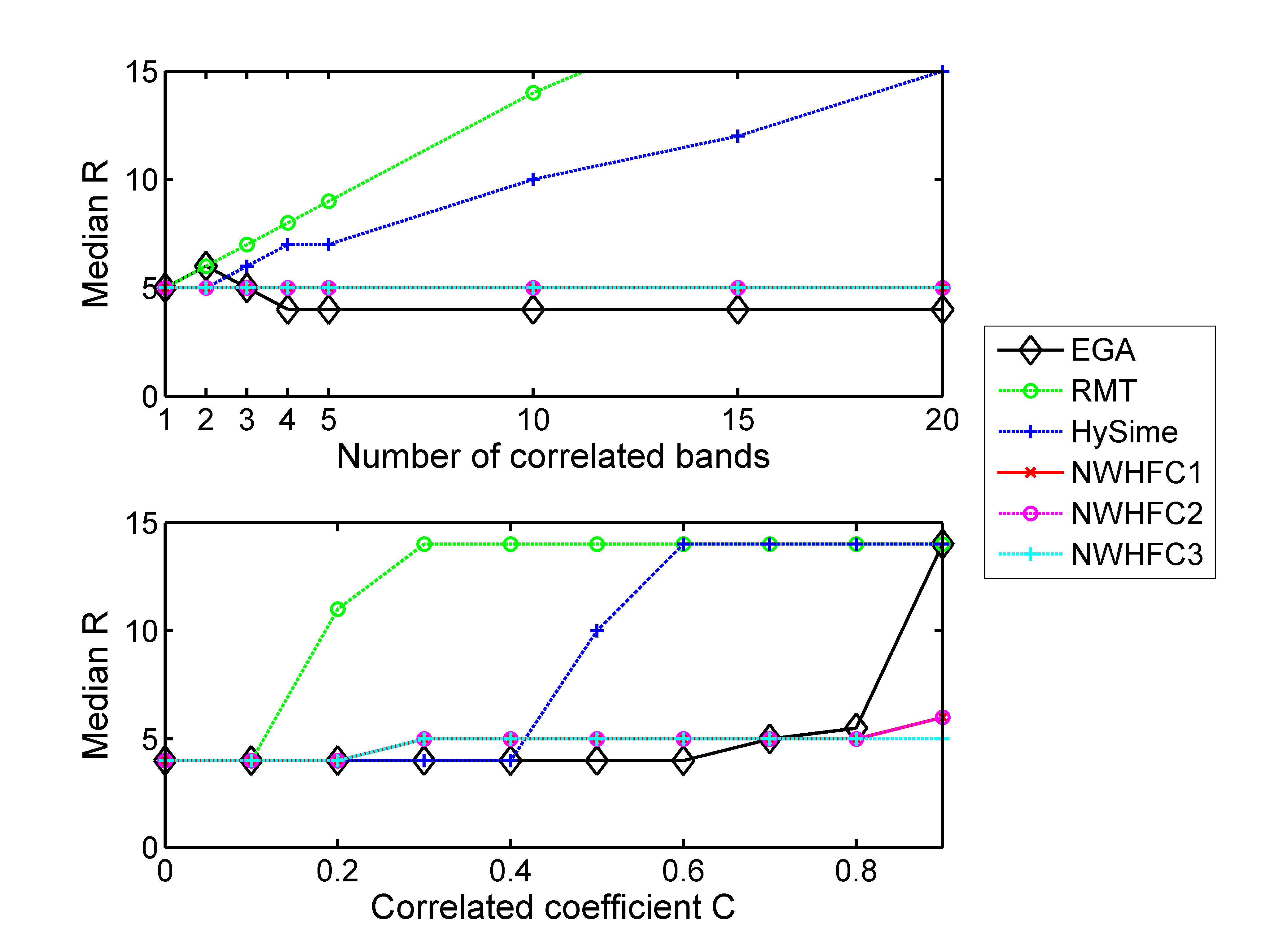}
\caption{Estimated R with respect to (top) number of correlated bands, (bottom) variation of the correlation coefficient. The actual number of endmembers is $R=4$.} \label{fig:Exp11_12_StudyCorr_Noise_50Simul}
\end{figure}

The second experiment considers effect of the noise correlation between the different spectral bands denoted as spectral correlation, that is generally observed in real data \cite{CawseWHISPERS2012,CawseJSTARS2013}.  To simulate data with spectral correlation, we considered the following covariance structure\footnote{We represented the covariance structure for one correlated band $j$. The case of multiple correlated bands can be obtained by considering multiple values for $j$.} when band $j$ is correlated with band $j+1$ with a correlation coefficient $C$
\begin{equation}
\bSig =    \left[ \begin{array}{cccccc}
   \sigma^2_1 &  0     &           & \cdots &  & 0 \\
  0           &  \ddots &           &  & & \vdots \\
	\vdots            &        & \sigma^2_j  & C\sigma^2_{j+1} &  & \\
	            &        & C\sigma^2_{j+1}  & \sigma^2_{j+1}&  & \\
	            &        &    &  & \ddots & \\
	0           &   0     &  \cdots  &  &  & \sigma^2_L
\end{array} \right].  \label{eqt:Cov_Correlation}
\end{equation}
This covariance structure was chosen to compare our results with  \cite{CawseJSTARS2013}, which used a similar matrix structure. We first varied the number of correlated spectral bands when considering a correlation coefficient $C=0.5$. The correlated bands are chosen randomly from the set $\left\lbrace1,\cdots,L-1\right\rbrace$. For all the algorithms, we   considered the noise estimation algorithm described in Section \ref{subsec:Noise_estimation}.  Fig. \ref{fig:Exp11_12_StudyCorr_Noise_50Simul} (top) shows a linear evolution of $\widehat{R}$  w.r.t.  the number of correlated bands for both RMT and Hysime (which is in agreement with the results of \cite{CawseJSTARS2013}). Both EGA and NWHFC show a stable result as the number of correlated bands increases. Note that EGA presents the best results. In a second study, we varied $C$ when considering $10$  correlated bands (drawn randomly between $1$ and $L$). The  results are shown in Fig. \ref{fig:Exp11_12_StudyCorr_Noise_50Simul} (bottom). The  EGA shows the best performance except for $C>0.8$ where NWHFC has a more stable results. To summarize, the obtained  results illustrate the robustness of the EGA with respect to noise estimation error and noise correlation. It is more robust to noise correlation than RMT, Hysime and NWHFC. Both EGA and NWHFC are robust to noise estimation error.

\subsection{Robustness to the image size} \label{subsec:Robustness_to_the_image_size}
As described in Section \ref{subsec:Rank_estimation_from_an_SPM}, the EGA is valid when $\gamma_1 > \cdots >\gamma_K > 1+\sqrt{c}$, with $c=\frac{L}{N}$. While this condition suggests that the image size should be large to obtain good results, we will see in this section that acceptable results are also obtained for small images.
The simulated images were obtained by using the previous $R=4$ endmembers and an i.i.d. Gaussian noise with $\textrm{SNR}=25$ dB.
Table \ref{tab:Perf_SizeN} shows the median of the estimated $\widehat{R}$ over $50$ Monte-Carlo results,  and the obtained accuracy indicated between brackets when varying the image size. All the algorithms provided poor results for $N=100$. However, both EGA and NWHFC provided accurate estimates for $N \geq 400$ pixels. Hysime offered accurate estimates for $N\geq 2500$ pixels while RMT required the largest number of pixels $N= 10000$ pixels. Note that the obtained results are in agreement with those of Section \ref{subsec:Robustness_to_noise}. Indeed, the estimated noise covariance $\widehat{\bSig}$ in \eqref{eq:CovEst_Noise} is sensitive to the number of pixels, that is, a reduced number of pixels   increases the estimation error of $\widehat{\bSig}$. Therefore, algorithms that are robust to noise estimates are expected to perform better when reducing the image size, which is observed in Table \ref{tab:Perf_SizeN}. To conclude, the results of this section show that EGA provides accurate results even for small images.
\begin{table}[h] \centering
\centering \caption{Estimated $R$ with respect to the image size $N$. Estimated median value and the accuracy in percent between brackets.}
\begin{tabular}{|c|c|c|c|c|c|}
\hline  Method   &  $N=10^2$  & $N=20^2$ & $N=30^2$ & $N=50^2$  & $N=10^4$ \\
\hline  EGA & 100 (0) &  4 (86) & 4 (100) &   4 (100)  &  4 (100) \\
  RMT        & 100 (0) &  63 (0) &  23 (0)  &   8 (0)  &   4 (100) \\
  HySime     & 100 (0) &  98 (0) &  29 (0)  &   4 (100) &  4 (100)\\
  $\textrm{NWHFC}_1$ & 67 (0) & 4 (100) & 4 (100) & 4 (100) & 4 (100)\\
  $\textrm{NWHFC}_2$ & 66 (0) & 4 (100) & 4 (100) & 4 (100) & 4 (100)\\
  $\textrm{NWHFC}_3$ & 66 (0) & 4 (100) & 4 (100) & 4 (100) & 4 (100)\\
\hline
\end{tabular}
\label{tab:Perf_SizeN}
\end{table}

\vspace{-1cm}
\subsection{Performance} \label{subsec:Performance}
This section studies the performance of the EGA when varying the number of endmembers, the noise level and the noise shape, as in \cite{Bioucas2008,AndreouJSTARS2014}.  The synthetic images were generated using the standard parameters described in Section \ref{sec:Simulation_results_on_synthetic_data}. For each Monte-Carlo simulation, the endmembers were randomly chosen in a database containing $20$ minerals. Moreover, and similarly to  \cite{Bioucas2008,AndreouJSTARS2014}, we considered two noise shapes w.r.t. spectral bands: (i) a constant shape w.r.t. spectral bands which represents an i.i.d. Gaussian noise and (ii) a Gaussian shape for the noise variance w.r.t. spectral bands defined as follows
\begin{equation}
\sigma^2_{\ell} = \sigma^2 \frac{\exp{\left[\frac{-(\ell-L/2)^2}{(2 \eta^2)}\right]}    }{\sum_i^L \exp{\left[\frac{-(i-L/2)^2}{(2 \eta^2)}\right]}}, \, \ell=1,\cdots,L \label{eqt:Gauss_Noise}
\end{equation}
where $\sigma^2$ is fixed according to the required SNR and $\eta$ controls the width of the Gaussian shape of the noise variance.
Table \ref{tab:Perf_NoiseIID} shows the obtained results with an  i.i.d. Gaussian noise. This table shows that all the algorithms provide good estimates for all SNRs when considering a reduced number of endmembers $R\leq 5$. However, the NWHFC algorithm shows poor results for large values of $R$ even for high SNRs. Note that Hysime, RMT and EGA algorithms provide good estimates for high SNR (SNR$>25$ dB) while the Hysime performance decreases for low SNR. Note finally that RMT and EGA provide similar performance.
Table \ref{tab:Perf_NoiseGauss} shows the results when considering a  Gaussian shape for the noise variance. This table shows poor results for NWHFC even for small values of $R$. However, the results are slightly improved when using the actual noise covariance matrix instead of the estimated one (see results between brackets). The Hysime, RMT and EGA algorithms  have a similar behavior as shown in Table \ref{tab:Perf_NoiseIID}, i.e., the Hysime performance decreases for low SNR while the RMT and EGA results are slightly better. These results show the accuracy of the EGA that provides equal or better results than the state-of-art algorithms.

\renewcommand{\arraystretch}{1.1}
\begin{table}[h] \centering
\centering \caption{Median of the estimated R for data corrupted by white noise (50 Monte Carlo simulations). For NWHFC, we show between brackets the results when using the ground-truth noise covariance matrix.}
\begin{tabular}{|c|c|c|c|c|c|}
\hline		  	SNR   &  Method  & R=3 & R = 5   & R = 10 & R = 15 \\
\hline	\multirow{6}{*}{15 dB} & EGA & \textbf{3} & \textbf{5} & 7 & \textbf{8} \\
	  & RMT    & 3 & 5 & \textbf{8} & 8 \\
	  & HySime & 3 & 4 & 5 & 4 \\
	  & $\textrm{NWHFC}_1$ & 3 (3) & 4 (4) & 3 (3) & 3 (3)\\
	  & $\textrm{NWHFC}_2$ & 3 (3) & 4 (4) & 3 (3) & 3 (3) \\
	  & $\textrm{NWHFC}_3$ & 3 (3) & 4 (4) & 3 (3) & 3 (3) \\
		
\hline	\multirow{6}{*}{25 dB} & EGA & \textbf{3}  &  \textbf{5}  & \textbf{10}  & \textbf{12} \\
	  & RMT    &  3  &  5  & 10  & 12 \\
	  & HySime &  3  &  5  &  8  &  9\\
	  & $\textrm{NWHFC}_1$ & 3 (3) & 4 (4) & 5 (5) & 5 (5)\\
	  & $\textrm{NWHFC}_2$ & 3 (3) & 4 (4) & 5 (5) & 5 (5) \\
	  & $\textrm{NWHFC}_3$ & 3 (3) & 4 (4) & 5 (5) & 4 (4) \\
		
\hline	\multirow{6}{*}{35 dB} & EGA & \textbf{3}  &  \textbf{5}  & \textbf{10}  & \textbf{15} \\
	  & RMT    & 3  &  5  & 10  & 15 \\
	  & HySime & 3  &  5  & 10  & 13 \\
	  & $\textrm{NWHFC}_1$ & 3 (3) & 4 (4) & 7 (7) & 7 (7)\\
	  & $\textrm{NWHFC}_2$ & 3 (3) & 4 (4) & 7 (7) & 6 (6)\\
	  & $\textrm{NWHFC}_3$ & 3 (3) & 4 (4) & 6 (6) & 6 (6)\\
		
\hline	\multirow{6}{*}{50 dB} & EGA & \textbf{3}  &  \textbf{5}  & \textbf{10}  & \textbf{15} \\
	  & RMT    & 3  &  5  & 10  & 15 \\
	  & HySime & 3  &  5  & 10  & 14 \\
	  & $\textrm{NWHFC}_1$ & 3 (3) & 4 (4) & 7 (7) & 9 (9)\\
	  & $\textrm{NWHFC}_2$ & 3 (3) & 4 (4) & 7 (7) & 8 (9)\\
	  & $\textrm{NWHFC}_3$ & 3 (3) & 4 (4) & 7 (7) & 8 (8)\\
\hline
\end{tabular}
\label{tab:Perf_NoiseIID}
\end{table}

\renewcommand{\arraystretch}{1.1}
\begin{table}[h] \centering
\centering \caption{Median of the estimated R for data corrupted by colored noise (Gaussian shape) with $50$ Monte Carlo simulations. For NWHFC, we show between brackets the results when using the ground-truth noise covariance matrix.}
\begin{tabular}{|c|c|c|c|c|c|}
\hline		  	SNR   &  Method  & R=3 & R = 5   & R = 10 & R = 15 \\
\hline	\multirow{6}{*}{15 dB} & EGA & \textbf{3}  &  \textbf{5}  &  6  &  6 \\
	  & RMT    & 3  &  5  &  5  &  5\\
	  & HySime & 3  &  4  &  5  &  5\\
	  & $\textrm{NWHFC}_1$ & 3 (3) & 5 (5) & \textbf{8} (\textbf{7}) & \textbf{9} (\textbf{9})\\
	  & $\textrm{NWHFC}_2$ & 3 (3) & 5 (5) & 8 (7) & 8 (8)\\
	  & $\textrm{NWHFC}_3$ & 3 (3) & 5 (4) & 7 (7) & 8 (8)\\
		
\hline	\multirow{6}{*}{25 dB} & EGA & \textbf{3}  &  \textbf{5}  &  \textbf{9}  & \textbf{10} \\
	  & RMT    & 3  &  5  &  8  &  9\\
	  & HySime & 3  &  5  &  8  &  8\\
	  & $\textrm{NWHFC}_1$ & 3 (3) & 5 (5) & 8 (7) & 9 (10)\\
	  & $\textrm{NWHFC}_2$ & 3 (3) & 5 (5) & 8 (7) & 9 (9)\\
	  & $\textrm{NWHFC}_3$ & 3 (3) & 5 (4) & 8 (7) & 8 (9)\\
		
\hline	\multirow{6}{*}{35 dB} & EGA & \textbf{3}  &  \textbf{5}  & \textbf{10}  & \textbf{14} \\
	  & RMT    & 3  &  5  & 10  & 13 \\
	  & HySime & 3  &  5  & 10  & 13 \\
	  & $\textrm{NWHFC}_1$ & 3 (3) & 5 (5) & 9 (7) & 11 (9)\\
	  & $\textrm{NWHFC}_2$ & 3 (3) & 5 (4) & 8 (7) & 10 (8)\\
	  & $\textrm{NWHFC}_3$ & 3 (3) & 5 (4) & 8 (7) & 9 (8)\\
		
\hline	\multirow{6}{*}{50 dB} & EGA & \textbf{3}  &  \textbf{5}  & \textbf{10}  & \textbf{15} \\
	  & RMT    & 3  &  5  & 10  & 15 \\
	  & HySime & 3  &  5  & 10  & 14 \\
	  & $\textrm{NWHFC}_1$ & 7 (3) & 9 (5) & 14 (7) & 18 (9)\\
	  & $\textrm{NWHFC}_2$ & 7 (3) & 9 (5) & 12 (7) & 16 (9)\\
	  & $\textrm{NWHFC}_3$ & 6 (3) & 7 (5) & 11 (7) & 15 (9)\\
\hline
\end{tabular}
\label{tab:Perf_NoiseGauss}
\end{table}


\section{Simulation results on real data} \label{sec:Simulation_results_on_real_data}
\begin{figure}[h!]
\centering \subfigure[]{\includegraphics[width=0.4\figwidth]{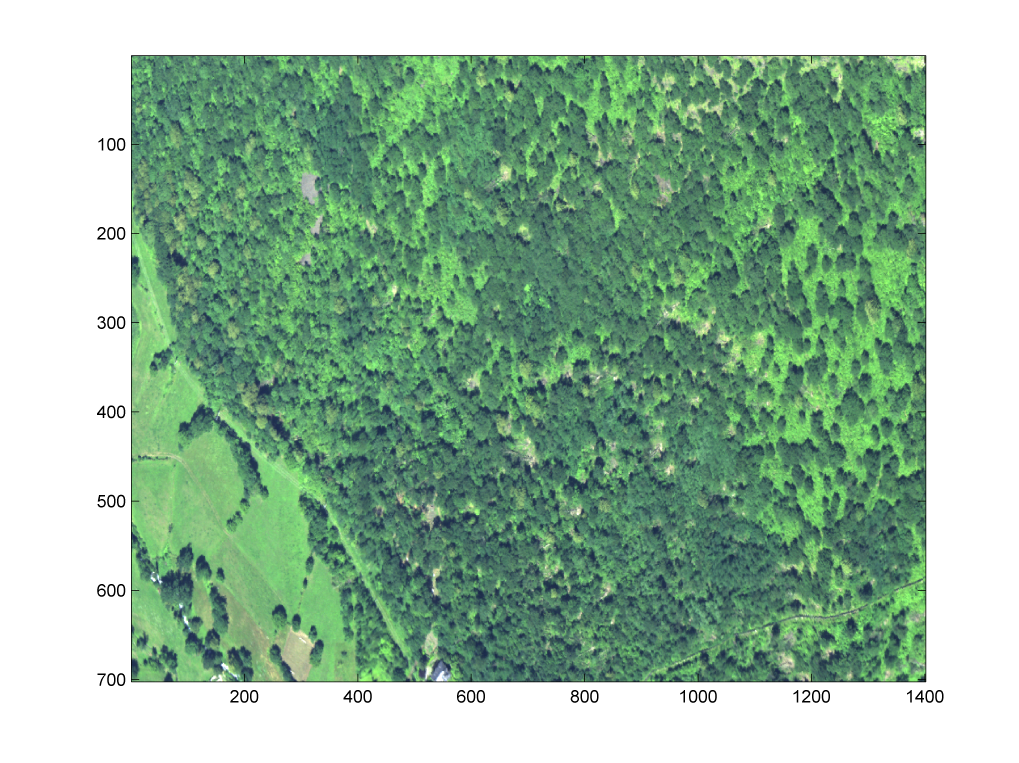}}
\subfigure[]{\includegraphics[width=0.4\figwidth]{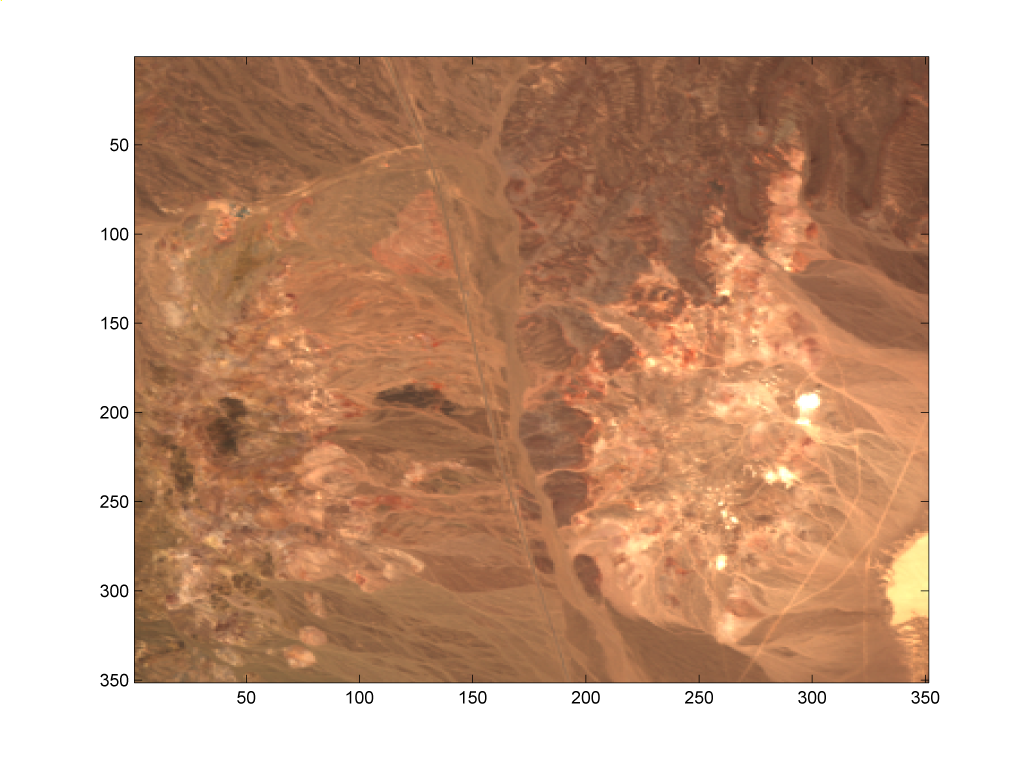}}
\subfigure[]{\includegraphics[width=0.4\figwidth]{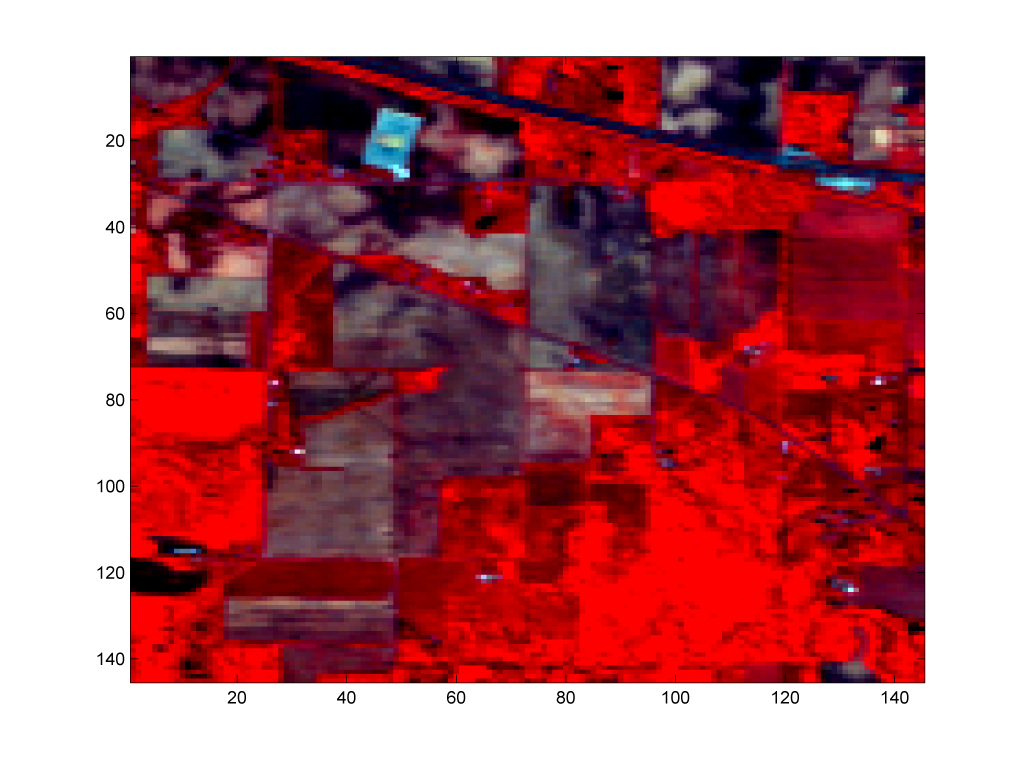}}
\caption{Real images. (a) Hyspex Madonna image, (b) AVIRIS Cuprite scene and (c) AVIRIS Indian pines  } \label{fig:Real_images}
\end{figure}

This section evaluates the EGA performance for three real hyperspectral images. The first  image was acquired in $2010$ by the Hyspex hyperspectral scanner over Villelongue, France (00 03'W and 4257'N). The dataset contains $L=160$ spectral bands recorded from the visible to near infrared ($400 - 1000$ nm)  with a spatial resolution of $0.5$ m  \cite{Sheeren2011}. The considered subset contains $702 \times 1401$ pixels and is mainly composed of forested areas \cite{Altmann2014b,HalimiArxiv2014} as shown in RGB colors in Fig. \ref{fig:Real_images} (a).  According to \cite{Sheeren2011}, the ground truth of this image contains $12$ tree species that are: ash tree, oak tree, hazel tree, locust tree, chestnut tree, lime tree, maple tree, beech tree, birch tree, willow tree, walnut tree and fern.  Consequently, the number of endmembers is expected to be at least equal to $12$. Table \ref{tab:Perf_Real_Im} (first column) shows the experimental results. The EGA estimated $R=12$ endmembers, which is in agreement with the ground truth information. The RMT, HFC and NWHFC provided a larger estimate while Hysime underestimated the number of endmembers. Note that the results obtained with HFC and NWHFC were expected since they estimate,  not only the endmember sources, but also the interferences \cite{Bioucas2012,AndreouJSTARS2014}.

The second image was acquired by the airborne visible/infrared
imaging spectrometer (AVIRIS) over the Cuprite mining site, Nevada, in $1997$. This image contains $182$ spectral bands with a spectral resolution of $10$ nm acquired in the 0.4-2.5 $\mu$m region (the water absorption bands $1$$-$$5$, $105$$-$$115$, $150$$-$$170$ and $220$$-$$224$ were removed) and a spatial resolution of $20$ m \cite{Kruze2002,CawseTIP2013}. The considered image subset contains  $351 \times 351$ pixels and is shown in RGB colors in Fig. \ref{fig:Real_images} (b). This image has been widely studied and a ground truth information is available. According to USGS\footnote{Available: \url{http://speclab.cr.usgs.gov/cuprite95.tgif.2.2um_map.gif}}, this image contains at least $18$ minerals \cite{Swayze1992}. The considered algorithms were applied to this image leading to the results in Table \ref{tab:Perf_Real_Im} (second column). All the algorithms estimated a   number of endmembers larger than $18$. EGA provided a more realistic value than RMT, which suffers from the spectral correlation when considering a multiple regression noise estimation algorithm \cite{CawseTIP2013}. However, the results obtained with Hysime, HFC and NWHFC were in better agreement with the ground truth (closer to $18$ endmembers).

The third   image was also   acquired  by the AVIRIS sensor, in june 1992 over an agricultural area of the northwestern Indiana\footnote{Available: \url{http://dynamo.ecn.purdue.edu/~biehl/MultiSpec/.}} (Indian Pines). The considered dataset contains $145 \times 145$ pixels, $185$ spectral bands with the same spectral resolution and spectral range as the Cuprite image (the water absorption bands $1$$-$$3$, $103$$-$$113$, $148$$-$$166$ and $221$$-$$224$ were removed) and a spatial resolution of $17$ m  \cite{Bioucas2008}. As shown in Fig. \ref{fig:Real_images} (c), the observed image is a mixture of agriculture and forestry.  According to the ground truth information \cite{Bioucas2008,AndreouJSTARS2014}, this image contains at least $16$ endmembers that are: alfalfa, corn-notill, corn-mintill, corn, grass-pasture, grass-trees, grass-pasture-mowed, hay-windrowed, oats, soybean-notill, soybean-mintill, soybean-clean, wheat, woods, buildings-grass-trees-drives and stone-steel-towers. Therefore, the estimated number should be greater than $16$. Table \ref{tab:Perf_Real_Im} (third column) reports the experimental results. Except Hysime that under-estimated the number of endmembers and HFC that over-estimated it, all algorithms detected $18$ components in the Indian Pines image.

The experimental results provided in this section illustrated the accuracy of the  EGA when applied to real data, acquired by different sensors (AVIRIS and Hyspex) and containing different physical elements (trees, grass and minerals).
%
%
\renewcommand{\arraystretch}{1.1}
\begin{table}[h] \centering
\centering \caption{Estimated $R$ for real images.}
\begin{tabular}{|c|c|c|c|}
\hline  Method       &  Madonna  & Cuprite & Indian Pines   \\
\hline  EGA         & 12 &  26 & 18  \\
				RMT          & 17 &  31 & 18    \\
				HySime       & 9  &  20 & 14   \\
  $\textrm{HFC}_1$   & 42 &  22 & 26  \\
  $\textrm{HFC}_2$   & 34 &  21 & 23  \\
  $\textrm{HFC}_3$   & 31 &  18 & 22  \\
  $\textrm{NWHFC}_1$ & 16 &  22 & 18  \\
  $\textrm{NWHFC}_2$ & 14 &  21 & 18  \\
  $\textrm{NWHFC}_3$ & 14 &  19 & 18  \\
\hline
\end{tabular}
\label{tab:Perf_Real_Im}
\end{table}\clearpage

\vspace{-1cm}
\section{Conclusions} \label{sec:Conclusions}
This paper proposed an unsupervised algorithm for determining the number of endmembers  in  hyperspectral images. This algorithm consisted  of two steps that are noise estimation and  determination of the endmember number. Noise estimation was achieved by a multiple regression estimation method even if other algorithms could be investigated. The second step was performed by thresholding the difference between successive eigenvalues of the sample covariance matrix. The resulting algorithm is non-parametric (it does not require any user-determined parameter) and efficient in the presence of i.i.d. and colored noise. Synthetic experiments showed a robust behavior of EGA with respect to noise estimation errors, noise correlations and noise levels. It also showed good performance when considering different image sizes. The obtained results on real images confirmed the accuracy of the proposed algorithm that showed comparable or better results than some state-of-the-art algorithms.
Future work includes the study of robust estimation for the pixel covariance matrix. Considering the recent method proposed in \cite{VinogradovaTSP2013,Vinogradova2013ICASSP} for source detection is also an interesting issue which would deserve to be investigated.

\newpage
\bibliographystyle{ieeetran}
\bibliography{biblio_all}

\begin{thebibliography}{10}
\providecommand{\url}[1]{#1}
\def\UrlFont{\rmfamily}
\providecommand{\newblock}{\relax}
\providecommand{\bibinfo}[2]{#2}
\providecommand\BIBentrySTDinterwordspacing{\spaceskip=0pt\relax}
\providecommand\BIBentryALTinterwordstretchfactor{4}
\providecommand\BIBentryALTinterwordspacing{\spaceskip=\fontdimen2\font plus
\BIBentryALTinterwordstretchfactor\fontdimen3\font minus
  \fontdimen4\font\relax}
\providecommand\BIBforeignlanguage[2]{{%
\expandafter\ifx\csname l@#1\endcsname\relax
\typeout{** WARNING: IEEEtran.bst: No hyphenation pattern has been}%
\typeout{** loaded for the language `#1'. Using the pattern for}%
\typeout{** the default language instead.}%
\else
\language=\csname l@#1\endcsname
\fi
#2}}

\bibitem{Keshava2002}
N.~Keshava and J.~F. Mustard, ``Spectral unmixing,'' \emph{IEEE Signal Process.
  Mag.}, vol.~19, no.~1, pp. 44--57, Jan. 2002.

\bibitem{Bioucas2012}
J.~Bioucas-Dias, A.~Plaza, N.~Dobigeon, M.~Parente, Q.~Du, P.~Gader, and
  J.~Chanussot, ``Hyperspectral unmixing overview: Geometrical, statistical,
  and sparse regression-based approaches,'' \emph{IEEE J. Sel. Topics Appl.
  Earth Observ. Remote Sens.}, vol.~5, no.~2, pp. 354--379, April 2012.

\bibitem{Dobigeon2014}
N.~Dobigeon, J.-Y. Tourneret, C.~Richard, J.~Bermudez, S.~McLaughlin, and
  A.~Hero, ``Nonlinear unmixing of hyperspectral images: Models and
  algorithms,'' \emph{IEEE Signal Process. Mag.}, vol.~31, no.~1, pp. 82--94,
  Jan 2014.

\bibitem{Halimi2011TGRS}
A.~Halimi, Y.~Altmann, N.~Dobigeon, and J.-Y. Tourneret, ``Nonlinear unmixing
  of hyperspectral images using a generalized bilinear model,'' \emph{IEEE
  Trans. Geosci. Remote Sens.}, vol.~49, no.~11, pp. 4153--4162, 2011.

\bibitem{AltmannTIP2011}
Y.~Altmann, A.~Halimi, N.~Dobigeon, and J.-Y. Tourneret, ``Supervised nonlinear
  spectral unmixing using a postnonlinear mixing model for hyperspectral
  imagery,'' \emph{IEEE Trans. Image Process.}, vol.~21, no.~6, pp. 3017--3025,
  June 2012.

\bibitem{Bioucas2008}
J.~M. {Bioucas-Dias} and J.~M.~P. Nascimento, ``Hyperspectral subspace
  identification,'' \emph{IEEE Trans. Geosci. Remote Sens.}, vol.~46, no.~8,
  pp. 2435--2445, Aug. 2008.

\bibitem{ChangTGRS2004}
C.~Chang and Q.~Du, ``Estimation of number of spectrally distinct signal
  sources in hyperspectral imagery,'' \emph{IEEE Trans. Geosci. Remote Sens.},
  vol.~42, no.~3, pp. 608--619, March 2004.

\bibitem{Nascimento2005}
J.~M. Nascimento and J.~M. {Bioucas-Dias}, ``Vertex component analysis: A fast
  algorithm to unmix hyperspectral data,'' \emph{IEEE Trans. Geosci. Remote
  Sens.}, vol.~43, no.~4, pp. 898--910, April 2005.

\bibitem{Dobigeon2009}
N.~Dobigeon, S.~Moussaoui, M.~Coulon, J.-Y. Tourneret, and A.~O. Hero, ``Joint
  {B}ayesian endmember extraction and linear unmixing for hyperspectral
  imagery,'' \emph{IEEE Trans. Signal Process.}, vol.~57, no.~11, pp.
  4355--4368, Nov. 2009.

\bibitem{Heinz2001}
D.~C. Heinz and {C. -I Chang}, ``Fully constrained least-squares linear
  spectral mixture analysis method for material quantification in hyperspectral
  imagery,'' \emph{IEEE Trans. Geosci. Remote Sens.}, vol.~29, no.~3, pp.
  529--545, March 2001.

\bibitem{Chen2014}
J.~Chen, C.~Richard, and P.~Honeine, ``Nonlinear estimation of material
  abundances in hyperspectral images with $\ell _{1}$-norm spatial
  regularization,'' \emph{IEEE Trans. Geosci. Remote Sens.}, vol.~52, no.~5,
  pp. 2654--2665, May 2014.

\bibitem{NguyenEAS2013}
N.~H. Nguyen, J.~Chen, C.~Richard, P.~Honeine, and C.~Theys, ``Supervised
  nonlinear unmixing of hyperspectral images using a pre-image methods,''
  \emph{EAS Publications Series}, vol.~59, pp. 417--437, Jan. 2013.

\bibitem{Ammanouil2014}
R.~Ammanouil, A.~Ferrari, C.~Richard, and D.~Mary, ``Blind and fully
  constrained unmixing of hyperspectral images,'' \emph{IEEE Trans. Image
  Process.}, vol.~23, no.~12, pp. 5510--5518, Dec 2014.

\bibitem{HalimiArxiv2014}
A.~Halimi, N.~Dobigeon, and J.-Y. Tourneret, ``Unsupervised unmixing of
  hyperspectral images accounting for endmember variability,'' in \emph{ArXiv
  e-prints}, Jun. 2014.

\bibitem{HoneineTGRS2012}
P.~Honeine and C.~Richard, ``Geometric unmixing of large hyperspectral images:
  a barycentric coordinate approach,'' \emph{IEEE Trans. Geosci. Remote Sens.},
  vol.~50, no.~6, pp. 2185--2195, June 2012.

\bibitem{Akaike1974}
H.~Akaike, ``A new look at the statistical model identification,'' \emph{IEEE
  Trans. Autom. Contr.}, vol.~19, pp. 716--723, 1974.

\bibitem{schwarz1978}
G.~Schwarz, ``Estimating the dimension of a model,'' \emph{The Annals of
  Statistics}, vol.~6, no.~2, pp. 461--464, 03 1978.

\bibitem{Rissanen1978}
J.~Rissanen, ``Modeling by shortest data description,'' \emph{Automatica},
  vol.~14, pp. 465--471, 1978.

\bibitem{Luo2013}
B.~Luo, J.~Chanussot, S.~Doute, and L.~Zhang, ``Empirical automatic estimation
  of the number of endmembers in hyperspectral images,'' \emph{IEEE Trans.
  Geosci. Remote Sens.}, vol.~10, no.~1, pp. 24--28, Jan 2013.

\bibitem{CawseTIP2013}
K.~Cawse-Nicholson, A.~B. Damelin, A.~Robin, and M.~Sears, ``Determining the
  intrinsic dimension of a hyperspectral image using random matrix theory,''
  \emph{IEEE Trans. Image Process.}, vol.~22, pp. 1301--1310, 2013.

\bibitem{CawseWHISPERS2011}
K.~Cawse, A.~Robin, and M.~Sears, ``The effect of noise whitening on methods
  for determining the intrinsic dimension of a hyperspectral image,'' in
  \emph{Proc. IEEE GRSS WHISPERS}, Lisbon, Portugal, June 2011, pp. 1--4.

\bibitem{CawseWHISPERS2012}
K.~Cawse-Nicholson, A.~Robin, and M.~Sears, ``The effect of spectrally
  correlated noise on noise estimation methods for hyperspectral images,'' in
  \emph{Proc. IEEE GRSS WHISPERS}, Shanghai, China, June 2012, pp. 1--4.

\bibitem{Passemier_RMTA12}
D.~Passemier and J.~F. Yao, ``On determining the number of spikes in a
  high-dimensional spiked population model,'' \emph{Random Matrices: Theory and
  Applications}, vol.~1, p.~19, 2012.

\bibitem{KritchmanTSP2009}
S.~Kritchman and B.~Nadler, ``Non-parametric detection of the number of
  signals: Hypothesis testing and random matrix theory,'' \emph{IEEE Trans.
  Signal Process.}, vol.~57, pp. 3930--3941, 2009.

\bibitem{Johnstone2001}
I.~M. Johnstone, ``On the distribution of the largest eigenvalue in principal
  component analysis,'' \emph{Ann. Stat.}, vol.~29, pp. 295--327, 2001.

\bibitem{Baik_JMA2006}
J.~Baik and J.~W. Silverstein, ``Eigenvalues of large sample covariance
  matrices of spiked population models,'' \emph{J. Multivariate Anal.},
  vol.~97, pp. 1382--1408, 2006.

\bibitem{PassemierPHD2012}
D.~Passemier, ``Inférence statistique dans un modèle à variance isolée de
  grande dimension,'' Ph.D. dissertation, Université Rennes 1, Rennes, France,
  2012.

\bibitem{Onatski_Econ08}
A.~Onatski, ``Testing hypotheses about the number of factors in large factors
  models,'' \emph{Econometrica}, vol.~77, pp. 1447--1479, 2009.

\bibitem{Altmann2014b}
Y.~Altmann, N.~Dobigeon, S.~McLaughlin, and J.-Y. Tourneret, ``Unsupervised
  post-nonlinear unmixing of hyperspectral images using a {H}amiltonian {M}onte
  {C}arlo algorithm,'' \emph{IEEE Trans. Image Process.}, vol.~23, no.~6, pp.
  2663--2675, June 2014.

\bibitem{Green_TGRS1988}
A.~Green, M.~Berman, P.~Switzer, and M.~Craig, ``A transformation for ordering
  multispectral data in terms of image quality with implications for noise
  removal,'' \emph{IEEE Trans. Geosci. Remote Sens.}, vol.~26, no.~1, pp.
  65--74, Jan 1988.

\bibitem{Meer_TPAMI1990}
P.~Meer, J.~Jolion, and A.~Rosenfeld, ``A fast parallel algorithm for blind
  estimation of noise variance,'' \emph{IEEE Trans. Pattern Anal. Mach.
  Intell.}, vol.~12, no.~2, pp. 216--223, Feb 1990.

\bibitem{RogerIGRS1996}
R.~E. Roger and J.~F. Arnold, ``Reliably estimating the noise in aviris
  hyperspectral images,'' \emph{Int. J. Remote Sens.}, vol.~17, no.~10, pp.
  1951--1962, 1996.

\bibitem{CawseJSTARS2013}
K.~Cawse-Nicholson, A.~Robin, and M.~Sears, ``The effect of correlation on
  determining the intrinsic dimension of a hyperspectral image,'' \emph{IEEE J.
  Sel. Topics Appl. Earth Observ. Remote Sens.}, vol.~6, no.~2, pp. 482--487,
  April 2013.

\bibitem{AndreouJSTARS2014}
C.~Andreou and V.~Karathanassi, ``Estimation of the number of endmembers using
  robust outlier detection method,'' \emph{IEEE J. Sel. Topics Appl. Earth
  Observ. Remote Sens.}, vol.~7, no.~1, pp. 247--256, Jan 2014.

\bibitem{Sheeren2011}
D.~Sheeren, M.~Fauvel, S.~Ladet, A.~Jacquin, G.~Bertoni, and A.~Gibon,
  ``Mapping ash tree colonization in an agricultural mountain landscape:
  {I}nvestigating the potential of hyperspectral imagery,'' in \emph{Proc. IEEE
  Int. Conf. Geosci. Remote Sens. (IGARSS)}, July 2011, pp. 3672--3675.

\bibitem{Kruze2002}
F.~A. Kruze, ``Comparison of {AVIRIS} and {H}yperion for hyperspectral mineral
  mapping,'' in \emph{{Proc. 11th JPL Airborne Geosci. Workshop}}, 2002, pp.
  1--11.

\bibitem{Swayze1992}
G.~Swayze, R.~Clark, S.~Sutley, and A.~Gallagher, ``Ground-truthing {AVIRIS}
  mineral mapping at {C}uprite, {N}evada,'' \emph{{Summaries 3 rd Annu. JPL
  Airborne Geosci. Workshop}}, vol.~1, pp. 47--49, 1992.

\bibitem{VinogradovaTSP2013}
J.~Vinogradova, R.~Couillet, and W.~Hachem, ``Statistical inference in large
  antenna arrays under unknown noise pattern,'' \emph{IEEE Trans. Signal
  Process.}, vol.~61, no.~22, pp. 5633--5645, Nov 2013.

\bibitem{Vinogradova2013ICASSP}
------, ``A new method for source detection, power estimation, and localization
  in large sensor networks under noise with unknown statistics,'' in
  \emph{Proc. IEEE Int. Conf. Acoust., Speech, and Signal Process. (ICASSP)},
  May 2013, pp. 3943--3946.

\end{thebibliography}
\end{document}